# Visualization of Cytoskeletal Elements by the Atomic Force Microscope

T. Berdyyeva[1], C.D. Woodworth[2], I. Sokolov[1]

[1]Dept. of Physics, [2]Dept. of Biology, Clarkson University, Potsdam, NY 13699, USA

## Abstract

We describe a novel application of atomic force microscopy (AFM) to directly visualize cytoskeletal fibers-in human foreskin epithelial cells. The nonionic detergent Triton X-100 in a low concentration was used to remove the membrane, soluble proteins, and organelles from the cell. The remaining cytoskeleton can then be directly visualized in either liquid or air-dried ambient conditions. These two types of scanning provide complimentary information. Scanning in liquid visualizes the surface filaments of the cytoskeleton, whereas scanning in air shows both the surface filaments and the total "volume" of the cytoskeletal fibers. The smallest fibers observed were ca. 50 nm in diameter. The lateral resolution of this technique was ca. 20 nm, which can be increased to a single nanometer level by choosing sharper AFM tips. Because the AFM is a true three dimensional technique, we are able to quantify the observed cytoskeleton by its density and volume. The types of fibers can be identified by their size, similar to electron microscopy.

# *INTRODUCTION*

Traditionally, transmission electron microscopy (TEM) and light microscopy are the techniques that have made major contributions to our understanding of cellular morphology. At present, there are three main techniques to study the cellular cytoskeleton: TEM [1,2], immunofluorescence microscopy [3,4], and developed recently, transmission X-ray microscopy [5]. It was found that the cytoskeleton consists of three types of filaments: microtubules, microfilaments, and intermediate filaments. Atomic force microscopy (AFM) is a newer technique, which has great potential in biology, primarily due to its ability to image biological objects directly in their natural conditions. Lateral resolution of the AFM on rigid surfaces can approach the atomic level. While scanning soft biological surfaces, the AFM can achieve a resolution of about 1nm [6-11]. The vertical resolution is mostly determined by the AFM scanner sensitivity, and typically is as high as 0.01nm. The AFM is broadly used to study cell morphology [12-20]. However, visualization of the cellular cytoskeleton has been restricted to fibers that are close to the cell surface [18,21-29].

In the present work, we describe a technique to visualize the cytoskeleton using AFM. We use the nonionic detergent Triton X-100 to remove the membrane, soluble proteins, and organelles from the cell. The concentration of the detergent is relatively low (0.5%) to prevent possible damage of the cytoskeleton. After removal of soluble proteins, the remaining cytoskeleton is imaged by the AFM in either contact or tapping modes. The advantages of this technique are the following: (1) One can quantitatively compare the cytoskeletons of different cells, for example, by measuring its volume per unit of projected area (see the Results and Discussion for more detail). (2) The detected signal does not depend of any staining agent, permeability of membrane for fluorescent dies, etc. (3) The imaging is non-destructive. There is no deterioration of the sample during the scanning. Samples can be easily preserved at 4°C or in dry gases for long term storage. (4) Ultrahigh vertical and single fiber lateral resolutions are easily reachable with a typical commercial AFM. (5) The technique requires a basic contact AFM mode. So, even the low-end AFM is suitable for the use of this method.

## MATERIALS AND METHODS

*Atomic Force Microscope.* Nanoscope Dimension 3100 by Digital Instruments/Veeco was used in the present study. The imaging was done in air and in liquid using the liquid fluid holder. A standard contact mode of scanning was used. All scanning in liquid was done in Hank's balanced salt solution (HBSS). A V-shaped standard narrow 200 μm AFM cantilever (Digital Instruments/Veeco, CA) was used through the study. The radius of the probe and its cleanness were tested on the reversed grid (TG01, Micromash, Inc.). A typical AFM tip used had the radius of the apex of ca.20-40 nm. The tip was cleaned before each series of measurements by a UV short-wave lamp for 2 min. The total signal on the AFM photodetector was set to 3-4V, whereas the negative setpoint 0.5 - 1V for imaging in liquid and 1 to 2 V while imaging in air. The scan rate was set *ca.* 2-3 Hz to optimize the image quality. Each image was collected in resolution of 512x512 pixels. The AFM tip was positioned over the cells with the help of built-in zoom optical system allowing observation areas from 150x110 to 675x510 μm$^2$ with 1.5μm resolution.

*Cells.* Primary cultures of human foreskin epithelial cells were prepared by a two-stage enzymatic digestion as described in [30] and cells were maintained in keratinocyte serum free medium (Invitrogen, Carlsbad, CA). The 60mm cell culture dishes were mounted on the chuck of the AFM with a double sticky tape.

*Cytoskeletal preparation for the imaging.* All protein removal was performed in 60 mm diameter plastic Petri dishes. The cells attached to the Petri dish were washed twice with HBSS solution. The cells were treated either for 10 min at room temperature or overnight at 4°C with a solution of 0.5% Triton X-100 detergent (Sigma) mixed with buffer [4] (10 mM M Tris–HCl, pH 7.6, 0.14 M NaCl, 5 mM MgCl$_2$, 4% polyethylene glycol 6000). After the treatment, the cells were washed twice for 2 min in the buffer and then fixed in the buffer with 1% formalin for 10 min. The treating solution was removed and the cells were washed twice with HBSS solution. The HBSS solution was added to scan the cell cytoskeleton in liquid. For scanning in air, the cells were washed with MilliQ ultrapure water and dried under ambient condition (24°C, 48% humidity). The cells were imaged immediately after preparation. However, the dry samples stored under nitrogen were imaged 2 days later without any traces of deterioration. Petri dishes with the cell cultures were mounted on the AFM chuck with a double sticky tape.

## *Results and Discussion*

### Influence and the choice of the treatment time

Imaging of viable cells by AFM is described in a host of literature. Although the cytoskeleton of viable cells is not directly visible some large fibers located close to the membrane can be observed. A typical unfiltered AFM scan of non-treated cells is shown in Fig.1. In contrast, cells treated with Triton X-100 in the way described in this work showed no traces of membrane. Moreover, the amount of soluble protein decreases over a few hours of treatment. This was detected by estimating the volume of the dried cells. When the treated cell dries, it collapses to the substrate. Consequently, the volume of the dried cell is approximately the volume of the dry proteins left after the treatment. A typical image of a treated cell while imaging in HBSS solution is shown in Fig.2a, and the same dried cell in Fig.2b. To be able to compare different cells quantitatively, we need to "normalize" that volume. By taking the ratio of the whole cell volume to the cell area, (i.e., calculating the volume per unit area), we can remove dependence on the cell lateral size. To make our estimation independent of the initial cell thickness, we calculate a coefficient of volume contraction, the ratio of the volume per unit area before and after the drying. The contraction coefficient will indicate the amount of protein left after the treatment. The larger the contraction coefficient due to drying, the smaller the amount of protein left. It is interesting to note that the total volume occupied by the cytoskeleton before drying did not change over the time of treatment. This implies that mechanically the cytoskeleton does not chance its morphology noticeably during the treatment. Averaged over five cells, the coefficient of contraction of the cells treated for 10 minutes (at room temperature) was ca. 3.7, whereas the contraction of the cells treated overnight (at $4^oC$ ) was ca. 5.7, and did not change for the extended treatment. This presumably means that all soluble parts were washed out. Based on this observation, we limited our study to cells treated overnight.

### Imaging in liquid vs. air dried

Let us comment on imaging in air versus imaging in liquids. Typically, it is harder to achieve good resolution while imaging in liquids. In contrast, the dried cytoskeletal structure is more rigid, and consequently, can be imaged by the AFM with higher resolution. Fig.3 shows an example of the same area on the cell imaged first (a) in HBSS solution, and (b) dried and imaged in air. The increase in resolution is clearly seen. The rest of images presented in this work are done in air.

It is also important to remember that the drying is associated with possible artifacts. Comparing the images of Fig.3, one can easily find the amount of the artifacts. As one can see for the example of Fig.3, these artifacts are not significant.

## Observation of Cytoskeletons of whole cells

Typical images of the cellular cytoskeleton of cells imaged in air are presented in Fig.4. In most cells, the nucleus (bright region) is surrounded by bright areas of fibers. The brightest, (i.e., the highest) spot which is seen in Fig.4 is the nucleolus. There are "white" bumps located close to the nuclei, which are presumably the components of the golgi or endoplasmic reticulum.
It should be noted that saturation in the nuclear region seen in Fig.4 is artificial, and it is done to stress the lower cytoskeleton fibers. Changing the vertical limit in the Nanoscope software (or tuning the contrast in any graphical editors), one can easily see the region of nucleus. Furthermore, the so-called deflection (sometimes called error signal) AFM image can be used to visualize small features while sacrificing the height information. Such an image can be treated as an analogy of Nomarksi differential imaging by a regular optical microscopy. Fig.5 shows one cell presented (a) with a small z-range to emphasize the cytoskeleton, (b) with a large z-range to show the height features around the nucleus, and (c) the deflection image of the cytoskeleton.

## High resolution imaging

Based on the observed fiber sizes and their position, we can conclude that we clearly image the fibers of F-actin between the cells in Figs.4 and 5. Each fiber image has the minimum observed lateral size of ca.50nm. This is partially a result of the AFM tip convolution [31]. Assuming the tip radius to be 30nm (average radius over the used tips range) and a cylindrical shape of the fiber, a simple geometrical solution gives us the deconvoluted diameter of the fiber of 20nm. The minimal vertical size of the filaments lying on the flat substrate near the cells is about 5-6 nm, which is close to the expected actin filament diameter of 6-10nm. Somewhat smaller observed values might be explained by deformation of the fibers by the AFM tip. Therefore, we can conclude that the observed fibers outside the cell consist of a bunch of actin filaments. Fig.6 shows a few such fibers. The arrow indicates the change of the left fiber's height from 12 to 6 nm. This presumably means that the fiber changes its height from two to one actin filament layer.

The lateral size of this fiber also changes from 135 nm (80 nm deconvoluted size) to 60 nm ( 22 nm deconvoluted size).

It should be noted that using the appropriate size of the AFM tip one can reach lateral resolution of a single nanometer level. For example, recently *ca.* 5nm resolution has been achieved in [32] while imaging microtubules with AFM.

## Quantitative measurements of cytoskeleton

Using the AFM images, we can quantify the cytoskeletal volume and its surface fiber density. To describe the fiber density we introduce a surface fiber density index (SFDI). The rationale behind introducing this index is as follows. The higher the density of the fibers, the smoother the cytoskeletal surface as imaged with the AFM. This is schematically shown in Fig.7. For example, the AFM image of the area of a very high density is almost flat. The difference between these two areas can then be treated as a measure of the density. To make it independent of the absolute value of the area, we will normalize it onto the flat (projected) area of the scan. To remove any dependence on large-scale surface corrugations, the AFM image should be flattened first. To make it unambiguous, we performed flattening of the $1^{st}$ order by using the "flatten" option in the Nanoscope software. This removes any linear tilt from each scan line of the AFM image. Finally, we can suggest the SFDI as given by the following formula:

$$SFDI = \frac{surface\ area - projected\ area}{projected\ area} 100\%$$

The smaller this index, the smoother the surface. Let us show how this formula works. Two cells presented in Fig.8 have rather different cytoskeletons. For example, the cell in Fig.8a, has a denser skeleton than the cell shown in Fig.8b. It also has a large amount of "bright bumps" presumably consisting of insoluble proteins that are components of the endoplasmic reticulum or lysosomes. Zoomed areas are shown in the same figure under those cells, Fig.8c,d. Each zoomed area has a line, along which the vertical section curve is shown below, Fig.8e,f. From these section curves, one can clearly see the difference in the fiber density. Using Nanoscope software (version 5.12r.4 roughness analysis option) to find the surface area, one can find the SFDI=0.70 (higher density) and =2.8 (lower density) for the cell of Fig.8a, and b, respectively. These values were calculated for the entire zoomed areas shown in Fig.8c,d. One can clearly see the correlation between the SFDI and the density of the cytoskeleton.

It should be noted that this index depends on the radius of the AFM tip used for the scanning. Additional study is required to make the SFDI index independent of the tip geometry. This will be done in the forthcoming study. For now, it can be used as a measure of a relative comparison utilizing the AFM tips of the same geometry.

To find the volume occupied by the cytoskeletal filaments, we used free software WSxM by Nanotec Electronica (version 4.0 develop 3.4). After drying, the cytoskeletal frame collapses to the surface. We assume that the volume of the collapsed structure is approximately equal to the volume of cytoskeletal fibers and insoluble components of the endoplasmic reticulum. To exclude contributions of the nucleus, the calculations of volume were done over the area away from the nuclear region. The volume was found by first removal of any overall tilt of the image, and then, by using option "flooding/find hills". To compare different cells, each volume was normalized by dividing the volume by the area it occupies.

Fig.9 shows an example of statistical variation of the calculated quantities over more than 20 cytoskeletons, the volume over area ratio and the SFDI index. A relatively large variation of the calculated quantities can probably be explained by various stages of cell development. This is in agreement with work described in Ref. [33] on the cytoskeleton of human peripheral blood lymphocytes, where it was showed that the concentration of F-actin fibers increases in older cells.

## *Conclusion*

We describe a relatively simple technique for direct visualization of cytoskeletal fibers by means of atomic force microscopy (AFM). Nonionic surfactant Triton X-100 in a low concentration was used to remove the membrane, soluble proteins, and organelles from normal human epithelial cells. The cytoskeletal filaments and non-soluble endoplasmic reticulum were imaged in both liquid and air-dried ambient conditions. The lateral resolution of this technique was ca.20 nm. Because the AFM is a true 3D technique, we are able to quantify the observed cytoskeleton by its density (the SFDI index) and volume (per unit area). The types of fibers can be identified by their size, similar to electron microscopy.

This new technique has the following advantages over existing techniques of TEM, XTM, or fluorescence microscopy. (1) The signal detected does not depend on any staining agent or permeability of the membrane for fluorescent dyes, which are used in all existing techniques. Therefore, one can qualitatively compare the cytoskeletons of different cells. (2) The imaging is non-destructive. There is no deterioration of the sample during the scanning like radiation damage in XTM and TEM, or fading of the florescent dye. Samples can be easily preserved at 4°C or in dry gases for a long storage, and reused later. (3) High lateral resolution is comparable with TEM, the best of existing techniques. Presumably, this resolution can even be increased because of the AFM's ability to reach molecular resolution on biological samples [6-11]. Specifically for microtubules, the resolution of ca. was attained 5nm in recent work [32]. The vertical resolution of presented here technique is well below the nanometer level, and surpasses any existing techniques. Both vertical and lateral resolutions are easily reachable with a typical commercial AFM. Finally, the technique requires only a basic contact AFM mode, so, the method works with even the low-end AFMs.

Disadvantages of the proposed technique can be outlined as follows. (1) Cells have to be treated. This excludes their scanning *in-vitro*. Despite almost every other technique requires to do the same type of destructive treatment, some optical fluorescent imaging allows obtaining the information about the cytoskeleton *in-vitro*. (2) The AFM is a surface analysis technique. Therefore, it is impossible to get the information about inside fibers. The other transmission techniques allow getting such information. (3) The treatment can potentially alter the cytoskeleton. Despite the treatment is close to the cell fixation used for XTM, at the present stage, it is not completely clear how the treatment can alter the cytoskeleton. Further study is needed to elaborate the suggested method.

To draw to a close, the suggested technique appears to have some advantages over the existing methods of studying the cytoskeleton. However, it clearly cannot provide all the information that may be obtained by the other methods. The suggested technique can be considered as a complimentary to the existing methods, giving in particular, rather accurate information about surface density of cytoskeletal fibers, and the total amount of fibers.

## *Acknowledgements*

I.S. is grateful to New York Center for Advanced Material processing (CAMP) and NSF (CCR-0304143) for partial support of this work.

## Figure Legend

Fig.1. An AFM image of a part of a viable cell. Only some large fibers located close to the membrane can be visualized.

Fig.2. 80 x80 µm$^2$ AFM scan of one cell treated for 10 minutes, imaged in (a) HBSS solution, and (b) the same cell imaged after drying in air. After drying the cytoskeleton and remaining proteins collapse to the substrate. (The brighter the area, the greater it's height.)

Fig.3. A 16x16 µm$^2$ AFM scan of an area of the cytoskeleton imaged in HBSS solution (right), and in air (left image).

Fig.4 The AFM images of the cellular cytoskeletons imaged in air. The bar size is 25 µm.

Fig.5. 77x77 µm$^2$ AFM images of the same cytoskeleton: (a) with a small z-range to emphasize the actin fibers, (b) with a large z-range to show the height features around the nucleus, (c) the deflection image of the cytoskeleton.

Fig.6. 3.5x3.5 µm$^2$ AFM image of actin fibers. The arrow indicates the change of the fiber height from 12 to 6 nm.

Fig.7. Lesser density fibers would produce higher corrugated surface image scanned with the AFM.

Fig.8. A 100x100 µm$^2$ AFM height scan of two different cytoskeletons, (a) and (b). The brighter means the higher. (c), (d) are 25x25 µm$^2$ zoomed areas of (a) and (b), respectively. Fig. (e) and (f) are the height profiles taken along the lines shown in (c) and (d) correspondingly. Once can see the higher density of the cytoskeleton in left figures, (a), (c), (e).

Fig.9. An example of statistical variation of the quantities, the volume/area ration and the SFDI index, calculated over more than 20 cytoskeletons.

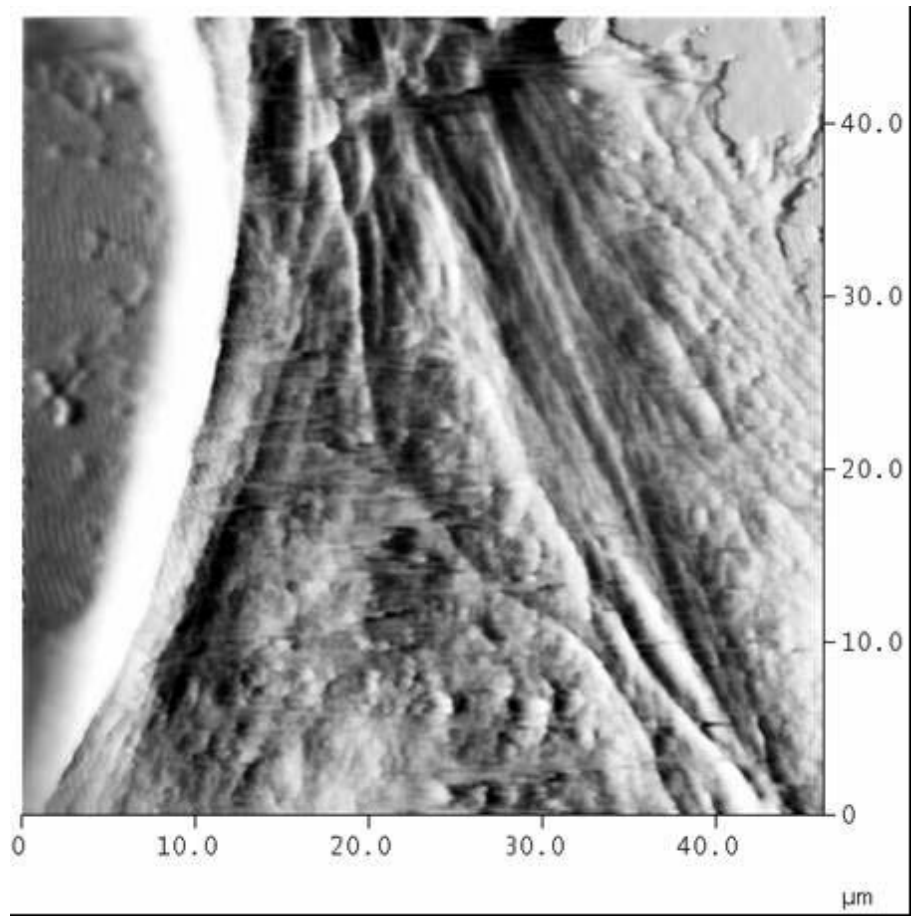

Fig.1

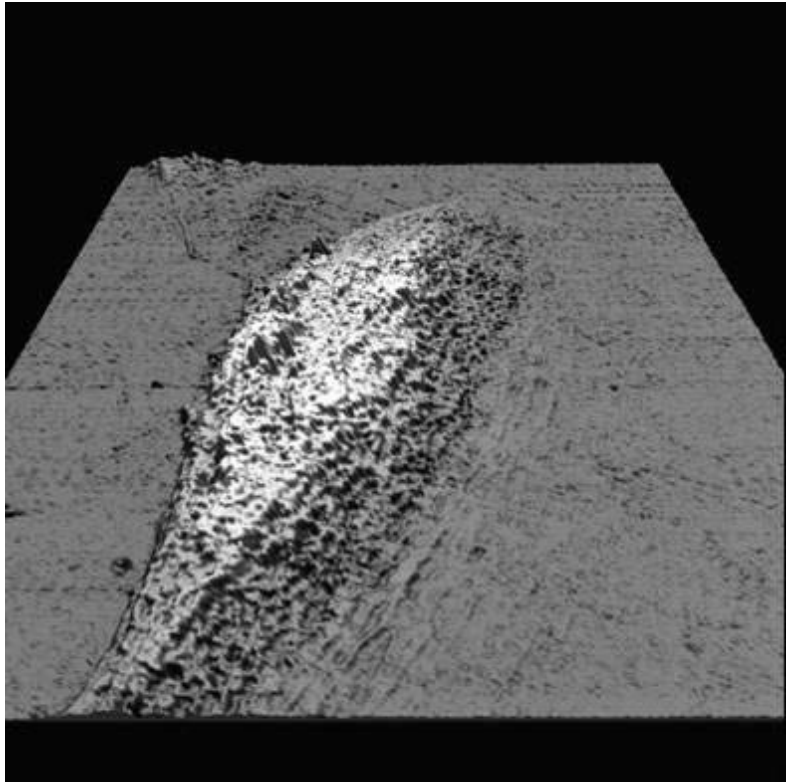

a

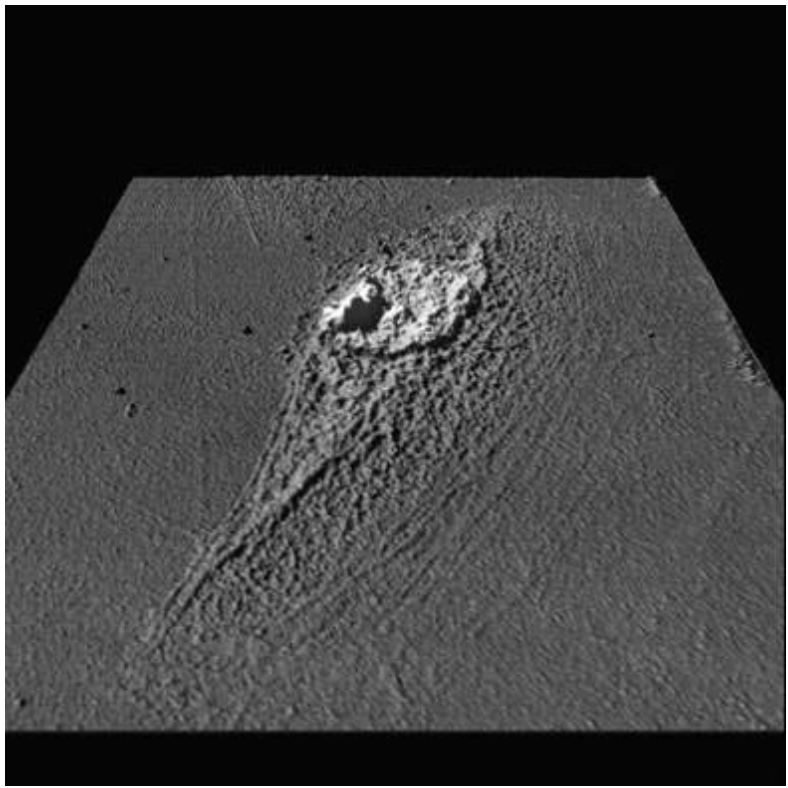

b

Fig.2

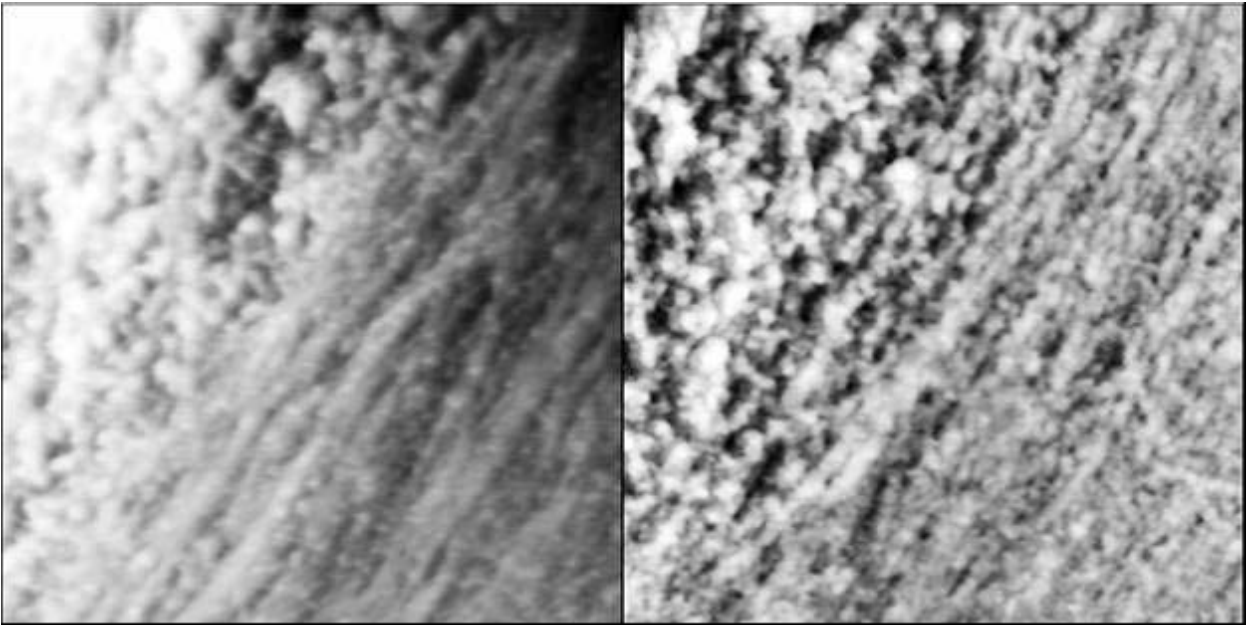

Fig.3

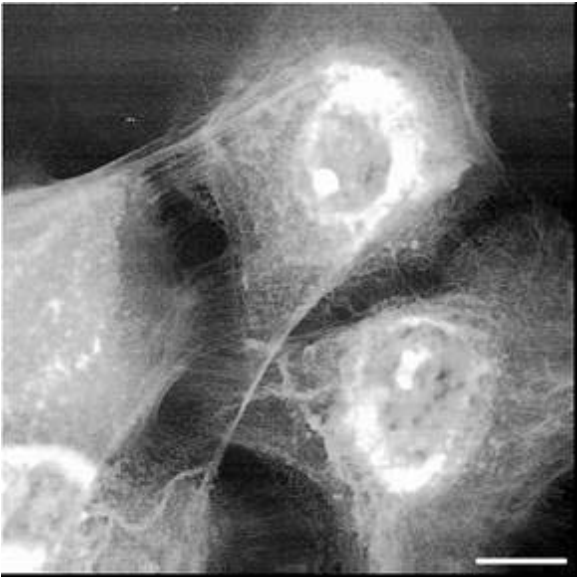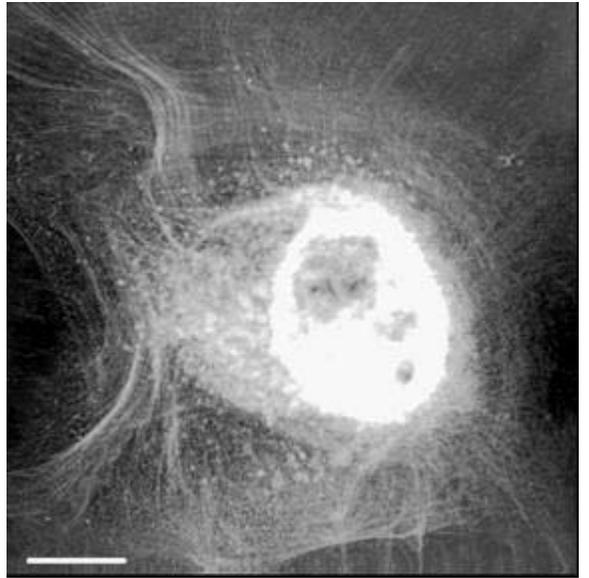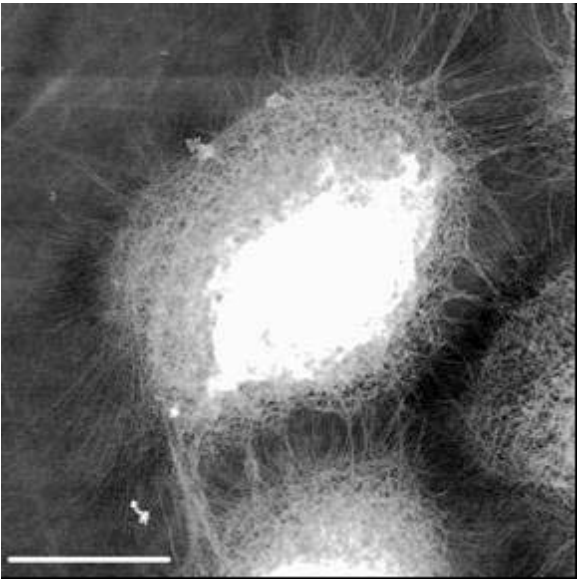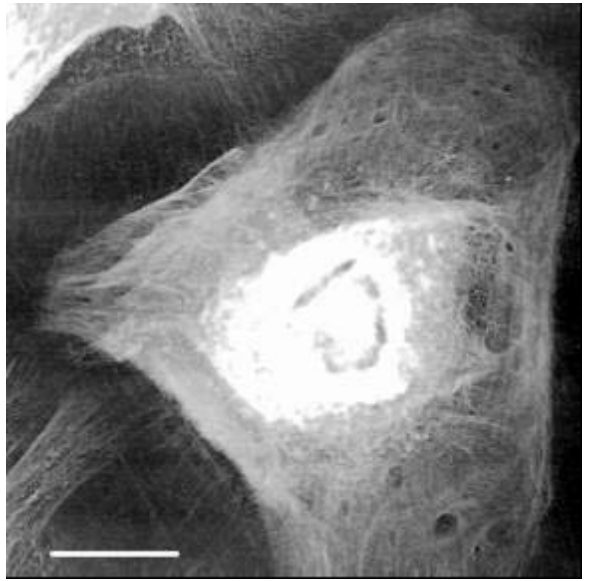

Fig.4

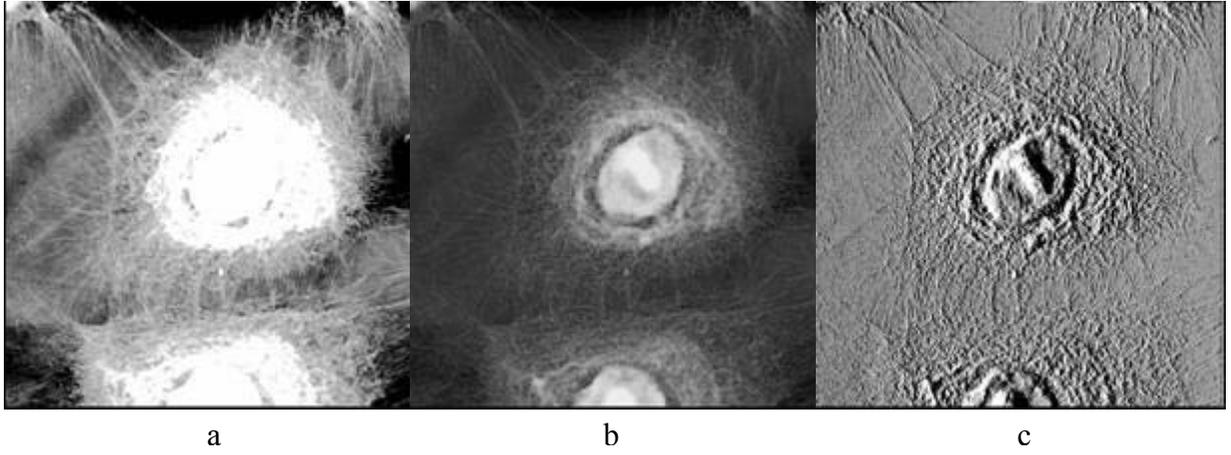

Fig.5

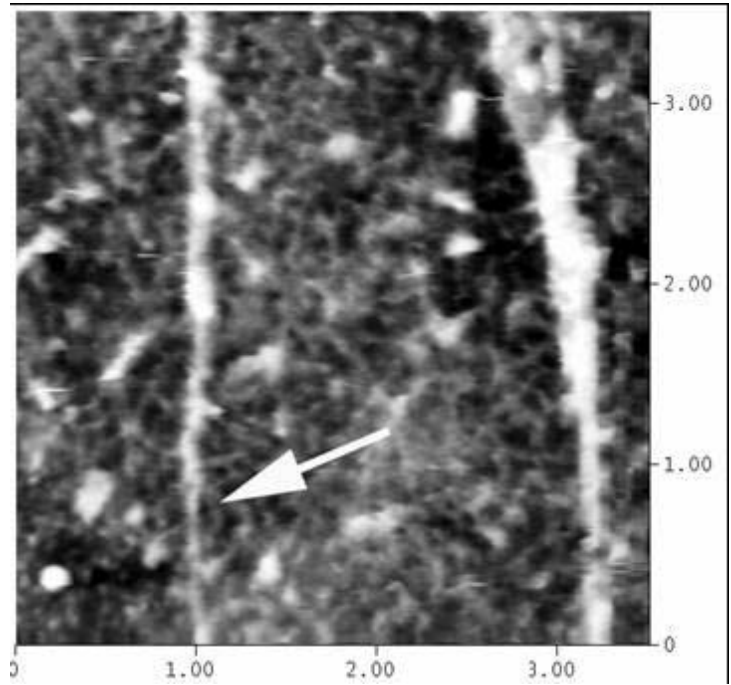

Fig.6

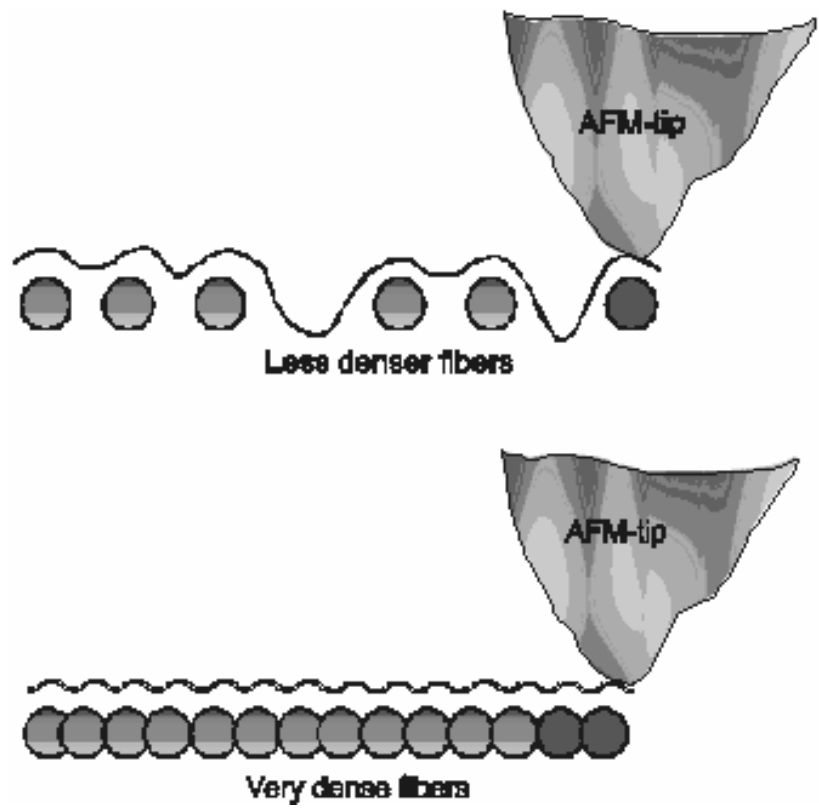

Fig.7

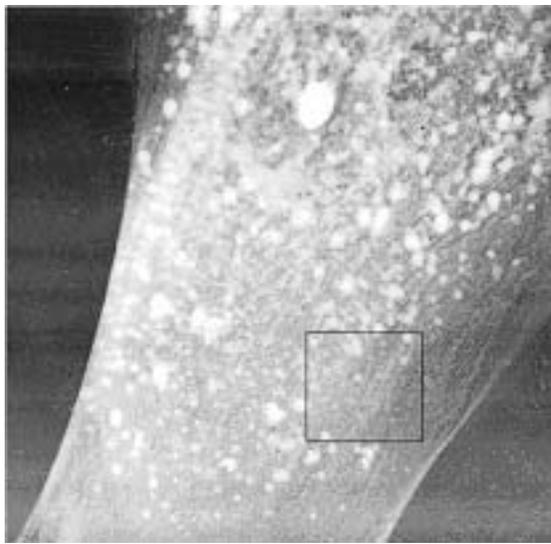
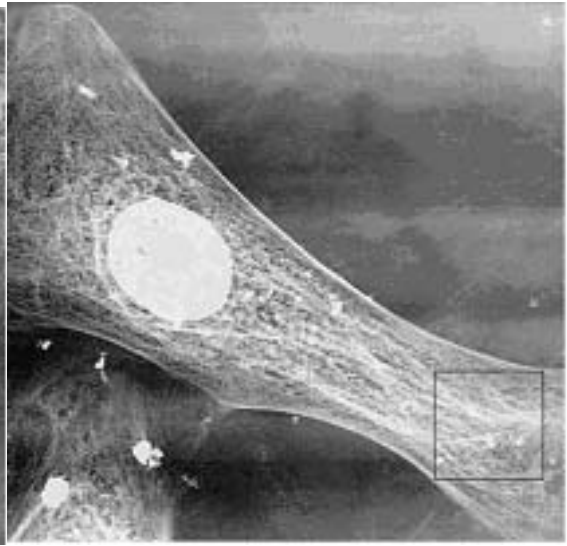

a  b

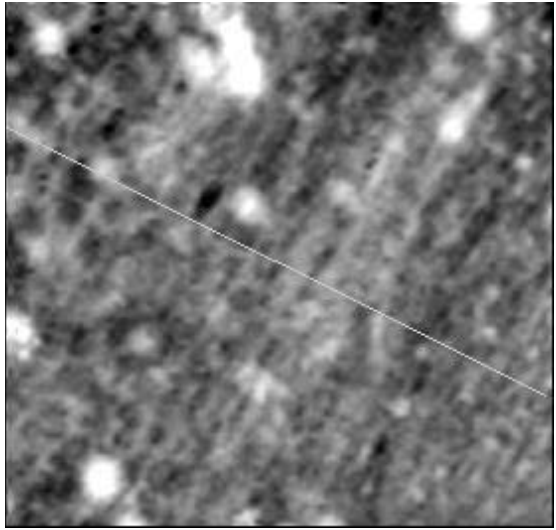
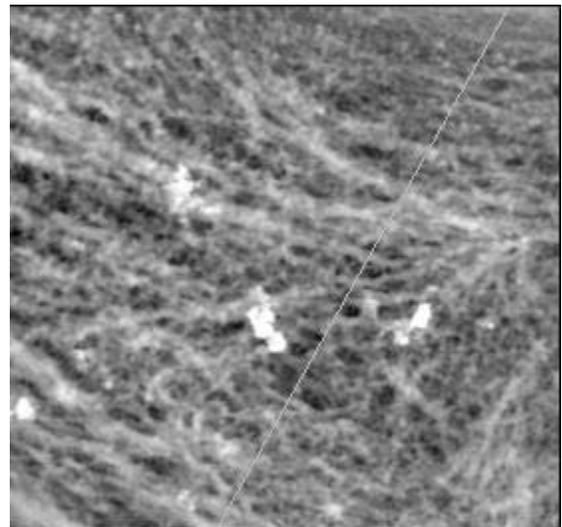

c  d

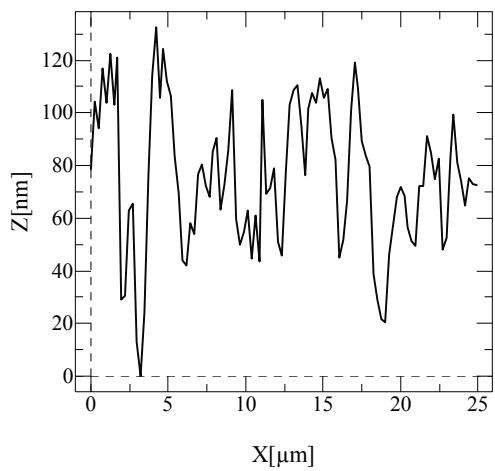
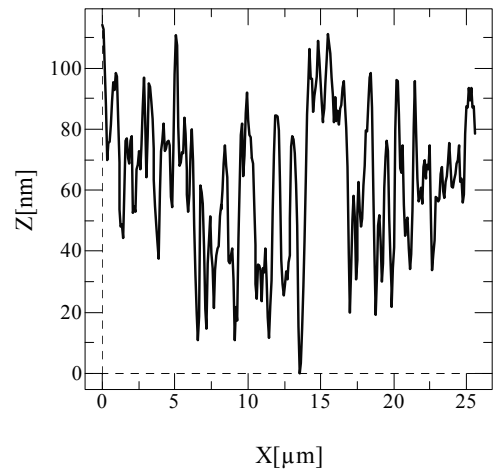

e  f

Fig.8

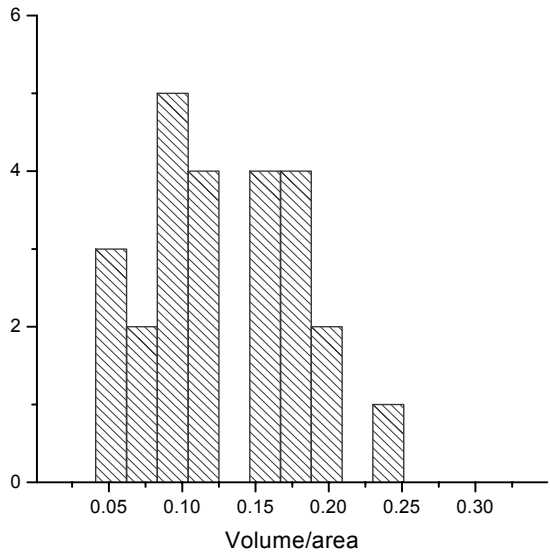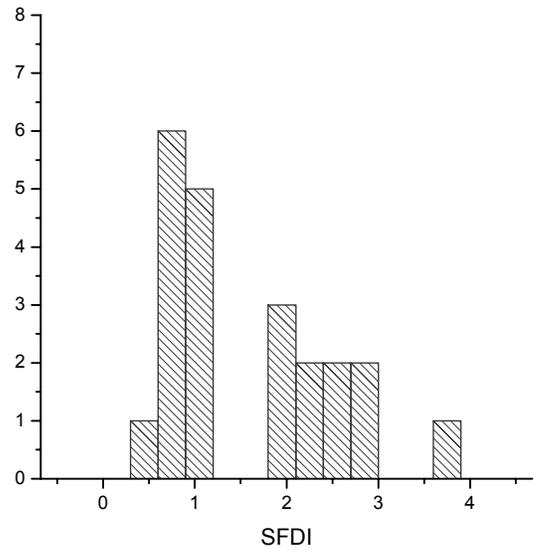

Fig.9